\documentclass{aa}

\usepackage{graphicx}
\usepackage{txfonts}
\usepackage{lipsum}
\usepackage{subcaption}
\usepackage{lscape}
\usepackage{placeins}
\usepackage{rotating}
\usepackage{soul}
\usepackage{float}

\begin{document}

   \title{Multivariate statistical analysis of low mass ratio contact binaries: definition, dynamical stability, and parameter relationships}
   \titlerunning{Statistical Analysis of Low Mass Ratio Contact Systems}

   \author{A. Poro\inst{1}
        \and R. Poggiani\inst{2}
        \and A. Foroutanfar\inst{3}
        \and R. Harzandjadidi\inst{4}
        \and N. Kahali Poor\inst{3}
        \and F. Alicavus\inst{5,6}
        }

   \institute{LUX, Observatoire de Paris, CNRS, PSL, 61 Avenue de l'Observatoire, 75014 Paris, France\\
             \email{atila.poro@obspm.fr}
            \and Department of Physics, University of Pisa, 56127 Pisa, Italy
            \and Independent Astrophysics Researcher, Tehran 15875, Iran
            \and Department of Theoretical Physics and Astrophysics, University of Zanjan, Zanjan, Iran
            \and Çanakkale Onsekiz Mart University, Faculty of Science, Department of Physics, 17020, Çanakkale, Türkiye\\
             \email{fahrialicavus@comu.edu.tr}
            \and Çanakkale Onsekiz Mart University, Astrophysics Research Center and Ulupnar Observatory, 17020, Çanakkale, Türkiye\\ }
   \date{Received ---, ---}

   \abstract
   {This study explores multiple aspects of W Ursae Majoris (W UMa) contact binary systems with low mass ratios, providing empirical insights into their definition, structure, rotational stability, and parameter relationships. We first examined the range of mass ratios that characterize these systems and, based on an analysis of 818 contact binaries, established an empirical threshold of $q \approx 0.27$ to identify low mass ratio systems. To investigate rotational stability, we conducted a Monte Carlo analysis of the squared gyration radii ($k_1^2$ and $k_2^2$) and assessed the resulting spin-to-orbital angular momentum ratio ($J_\mathrm{spin}/J_\mathrm{orb}$), finding that while $k_1$ remains nearly constant, $k_2$ and $J_\mathrm{spin}/J_\mathrm{orb}$ decrease slightly with increasing mass ratio, emphasizing the role of the secondary star's internal structure. Moreover, we compiled a dedicated sample of 115 low mass ratio contact binaries and estimated their absolute parameters using Gaia DR3 parallaxes. From this dataset, we derived empirical parameter relationships for low mass ratio systems, which provide a useful reference for future observational and theoretical studies. The resulting datasets and statistical summaries offer benchmarks for modeling, stability evaluation, and evolutionary studies of W UMa-type binaries with low mass ratios.}


   \maketitle
   \nolinenumbers

\section{Introduction}
Contact binaries are close binary systems with short orbital periods in which both stars fill their Roche lobes and share a common convective envelope (\citealt{Eggleton2006}). The energy generated by the components is transported throughout this envelope, producing an almost uniform surface temperature that does not strongly depend on the stellar masses (\citealt{1968ApJ...151.1123L,1981ARA&A..19..277S, Eggleton2006}). At sufficiently high orbital inclinations, these systems are classified as W UMa type eclipsing binaries. They are more likely to develop starspots than other types of close binaries, due to mass and energy transfer as well as magnetic activity (\citealt{1951PRCO....2...85O,1990ApJ...355..271Z}).

An important parameter of contact binaries is the mass ratio, $q = M_{2}/M_{1}$ ($q \leq 1$). After formation, contact binaries evolve toward lower mass ratio values in a nonlinear way (\citealt{2023A&A...672A.176P}). They may undergo Thermal Relaxation Oscillations (TROs), during which mass transfer can briefly reverse and contact is broken (\citealt{1976ApJ...205..208L, 1977MNRAS.179..359R, 2005ApJ...629.1055Y}). The TRO cycle depends on the secondary's thermal timescale, which grows as mass ratio decreases, causing systems to accumulate at low mass ratio (\citealt{2001AJ....122.1007R}). Alternative models suggest rapid initial mass transfer followed by mass ratio inversion and roughly linear evolution toward low mass ratio (\citealt{2011AcA....61..139S}). Mass inequality grows until tidal Darwin instability occurs, when the primary's spin exceeds one-third of the orbital angular momentum (\citealt{1980A&A....92..167H}). Previous studies have suggested that contact binaries with extremely low mass ratios may reach tidal instability, as described by Darwin, which can lead to rapid merger events (\citealt{2006MNRAS.369.2001L, 2010MNRAS.405.2485J,2026ApJ...P}). However, the precise value of this lower mass ratio limit remains a subject of debate. Theoretical predictions have placed this cutoff within a broad range, typically between $q \sim 0.04$ and $q \sim 0.10$, depending on assumptions regarding stellar structure, angular momentum distribution, and evolutionary stage (\citealt{2007MNRAS.377.1635A, 2009MNRAS.394..501A, 2015AJ....150...69Y, 2024SerAJ.208....1A}).

Observationally, several overcontact binaries with $q < 0.1$ have been reported (e.g., \citealt{2015AJ....150...69Y, 2024A&A...692L...4L}), challenging the strict theoretical limits and suggesting that at least some very low mass ratio systems may persist in a quasi-stable state. These findings indicate that the minimum mass ratio may not represent an absolute cutoff, but rather a transitional range influenced by system-specific conditions such as total mass, fillout factor, and angular momentum loss.

Despite extensive theoretical and observational efforts, the fundamental processes governing the formation, evolution, and stability of contact binaries with low mass ratios remain poorly understood. Current models disagree on whether systems with extremely low mass ratios can maintain long-term dynamical stability or inevitably evolve toward merger, and observational constraints are further limited by selection effects, small sample sizes, and systematic uncertainties in determining the mass ratio (\citealt{2006MNRAS.369.2001L,2007MNRAS.377.1635A}). Moreover, the empirical boundary separating stable low mass ratio systems from those approaching tidal instability has not yet been reliably established, partly due to the absence of uniform analyses across large, homogeneous datasets (\citealt{2001AJ....122.1007R}). Defining this boundary is essential for understanding the ultimate fate of contact binaries, their connection to binary-merger pathways, and their contribution to the formation of blue stragglers and FK Com-type stars (\citealt{1976ApJ...209..829W}). These gaps highlight the need for a comprehensive statistical investigation based on updated observational samples and consistent modeling. The structure of the paper is as follows: Section 2 presents the results of our empirical study on defining low mass ratios in contact binary systems; Section 3 presents the evaluation of squared gyration radii and their impact on spin-orbit stability in low mass ratio contact binaries; Section 4 examines the empirical parameters relationships for contact binaries with low mass ratios; and Section 5 provides the discussion and conclusions.

\section{Definition of the low mass ratio range}
Numerous studies have investigated the properties of W UMa-type contact binaries, focusing on how their structural and dynamical characteristics vary with system parameters (e.g., \citealt{2001AJ....122.1007R, 2001MNRAS.328..635Q, 2020MNRAS.497.3493Z, 2022MNRAS.510.5315P, 2023A&A...672A.176P}). Many studies consider systems with $q \lesssim 0.25$ as belonging to the low mass ratio range, although this range has not been established through systematic statistical or empirical analysis. In the literature, the definition of a low mass ratio in contact binaries varies among studies. Accordingly, one of the objectives of this study is to more precisely constrain the range of low mass ratios in contact binary systems using statistical and empirical analyses.

\begin{figure}
 \centering
 \includegraphics[width=0.5\textwidth]{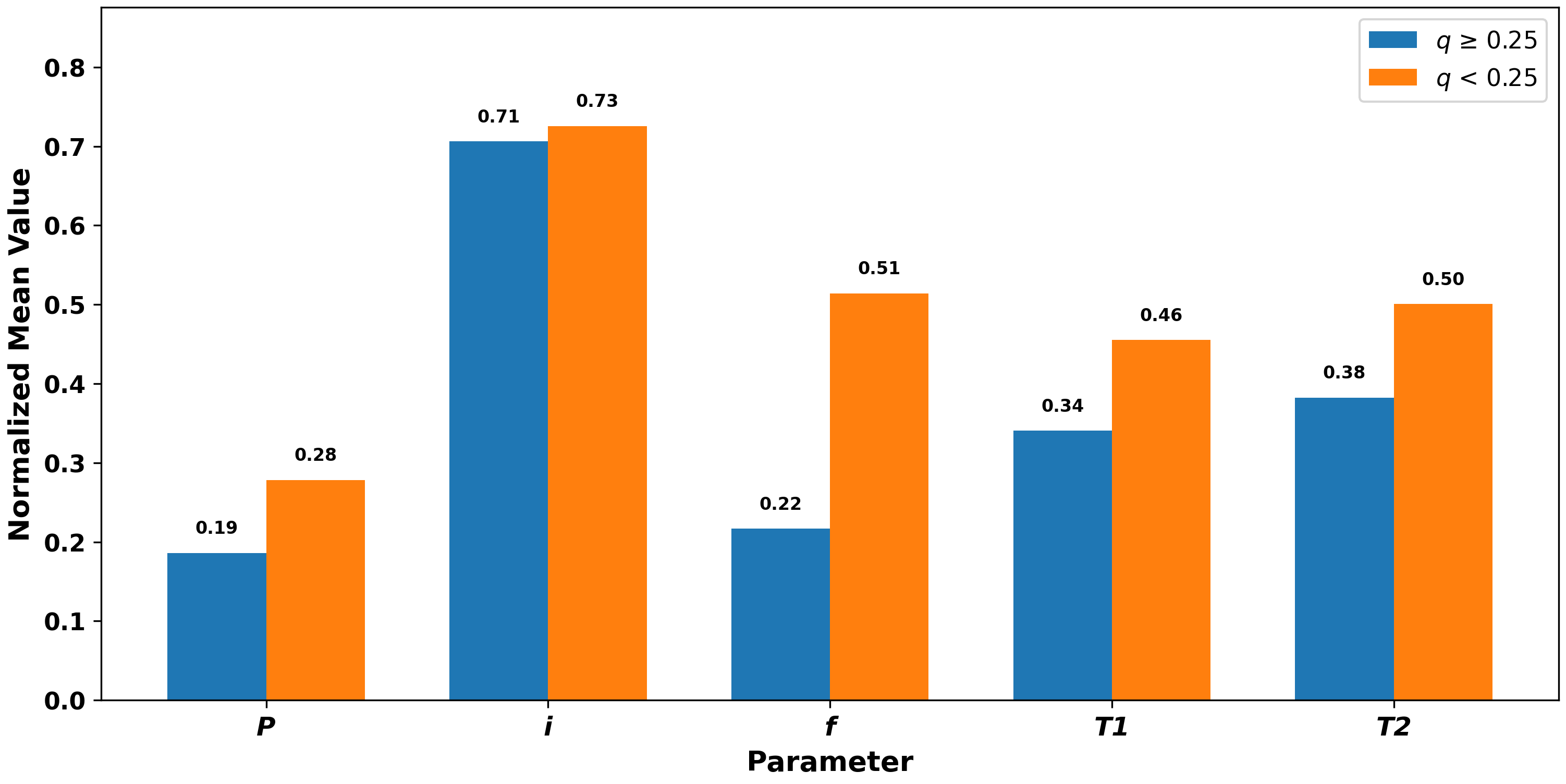}
 \caption{Comparison of normalized mean values of key parameters between contact binary systems with low mass ratios ($q < 0.25$, orange) and higher mass ratios ($q \ge 0.25$, blue). Each parameter is scaled to [0, 1] to allow direct comparison.}
 \label{fig:normalized_comparison}
\end{figure}

We analyzed a sample of 818 contact binary systems compiled by \cite{2025MNRAS.538.1427P} in order to determine the parameter range of low mass ratio systems. The analysis focused on a set of independent parameters derived from light curve modeling and observational data: the orbital period ($P$), mass ratio ($q$), fillout factor ($f$), and component temperatures ($T_{1,2}$). Parameters such as the luminosity ratio ($l_1/l_2$) and the fractional radii ratio ($r_1/r_2$), which depend directly on $q$, were not treated as independent variables and were therefore excluded from this study. Based on assumptions from previous studies, a threshold mass ratio of $q = 0.25$ was adopted to separate low mass ratio systems from higher-mass ratio systems. This criterion was then used to compare the distributions of independent parameters between systems with $q < 0.25$ and those with $q \geq 0.25$. Our analysis, based on a statistical comparison of the independent parameters between systems with $q < 0.25$ and $q \geq 0.25$, including the use of the Mann-Whitney U test to assess the significance of the differences and a decision tree to evaluate feature importance, shows that the fillout factor exhibits the largest difference across this threshold. Systems with $q < 0.25$ typically exhibit a broad distribution of fillout factors, with their average degree of contact tending to be higher. Conversely, systems with $q \geq 0.25$ predominantly show lower degrees of contact. This trend may reflect the underlying binary interaction processes. Low mass ratio systems are more likely to evolve into deeper contact configurations due to the combined effects of Roche geometry, mass transfer and angular momentum loss. Energy and mass redistribution within the overcontact region also contributes to a higher fillout factor in these systems. The orbital period exhibits the next largest variation, with low mass ratio systems generally having shorter periods, as expected for systems in which the components are more closely packed to maintain gravitational equilibrium under high overcontact conditions. Other parameters, such as component temperatures, show smaller systematic differences across the threshold. In summary, the feature importance analysis indicates that the fillout factor contributes the most to distinguishing low- and high-mass ratio systems, with 39.6\% importance, followed by the orbital period with 18.7\%, secondary temperature with 15.9\%, and the primary temperature with 14.4\% (Figure \ref{fig:normalized_comparison}). These results are based on the application of Min-Max Scaling, a common method for normalizing mean values (Figure \ref{fig:normalized_comparison}). This normalization enables comparison of parameters within a similar range. Additionally, we examined the Standard RobustScaler and Absolute RobustScaler methods. Ultimately, the final results indicated no significant differences among the scaling techniques.

\begin{figure}
\centering
\includegraphics[width=0.5\textwidth]{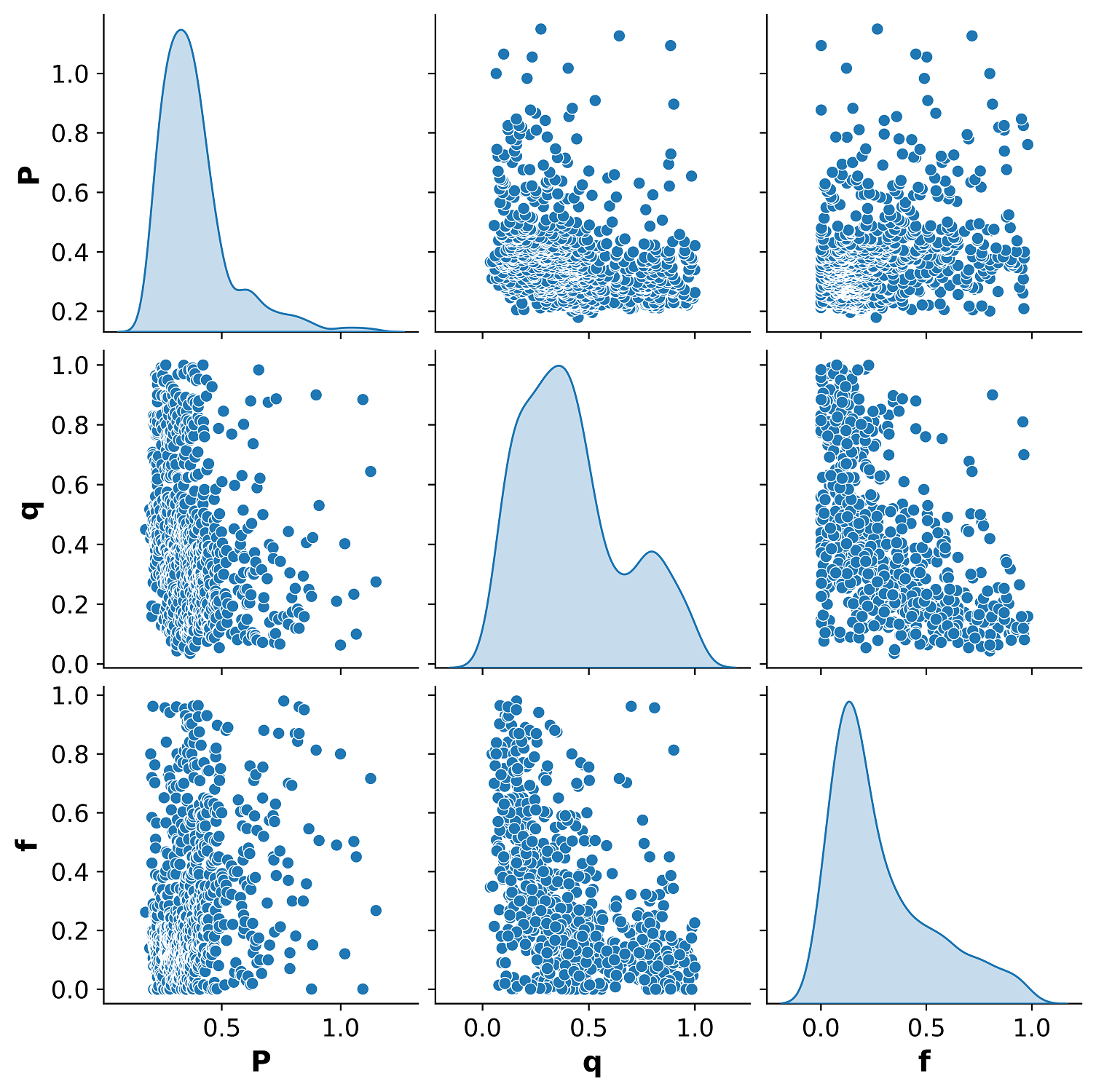}
\caption{Pairwise scatter plots of orbital period, mass ratio, and fillout factor for the contact binary sample, with KDE on the diagonals.}
\label{Fig:corl}
\end{figure}

Then, we examined the pairwise correlations among the orbital period, mass ratio, and fillout factor. The correlation analysis, expressed in terms of the Pearson correlation coefficient $r$, shows that the orbital period is weakly anti-correlated with the mass ratio ($r = -0.27$) and weakly positively correlated with the fillout factor ($r = 0.28$), while the mass ratio and the fillout factor exhibit a moderate negative correlation ($r = -0.50$). These results indicate that no two parameters alone display a strong linear dependence. Nevertheless, when all three parameters are considered together, low mass ratio systems can be identified more clearly than using any two parameters alone. Figure \ref{Fig:corl} shows the pairwise scatter plots of $P$, $q$, and $f$, with kernel density estimation (KDE) applied on the diagonals to visualize the distributions. The mass ratio distribution appears to be double-peaked in Figure \ref{Fig:corl}, with a clear maximum around \( q \sim 0.4 \), while the second peak is less pronounced.

To investigate the three-way interdependence of these parameters and define a statistical range for low mass ratio contact systems, we applied a three-dimensional KDE approach. KDE provides a smooth, continuous estimate of the probability density function in the joint parameter space defined by $\log P$, $f$, and $q$. For each point in this space, the density is estimated as the weighted contribution of all surrounding systems, where the weighting is controlled by a kernel function (typically Gaussian) and a bandwidth parameter that sets the effective smoothing scale. By evaluating this density field, it is possible to identify regions of high statistical concentration corresponding to physically common system configurations, as well as sharp declines in density that suggest natural empirical boundaries.

In our analysis, we projected the KDE-derived density distribution onto the $q$ axis while preserving the influence of $\log P$ and $f$. This projection reveals how the overall density of contact binaries changes as a function of mass ratio, conditioned by the other two parameters. The distribution exhibits a marked drop around $q \approx 0.27$, below which the density of systems declines significantly. This transition is not imposed by an arbitrary cutoff but emerges naturally from the underlying density field. Systems with $q \lesssim 0.27$ were therefore identified as low mass ratio candidates, with this threshold representing the point at which the KDE-estimated density falls below a statistically significant level relative to the surrounding distribution. This threshold corresponds to a statistically significant level, which refers to a density drop that is more than 2-3 standard deviations lower than the surrounding data, indicating that the observed difference is unlikely to be due to random effects.

\begin{figure*}
\centering
\includegraphics[width=1\textwidth]{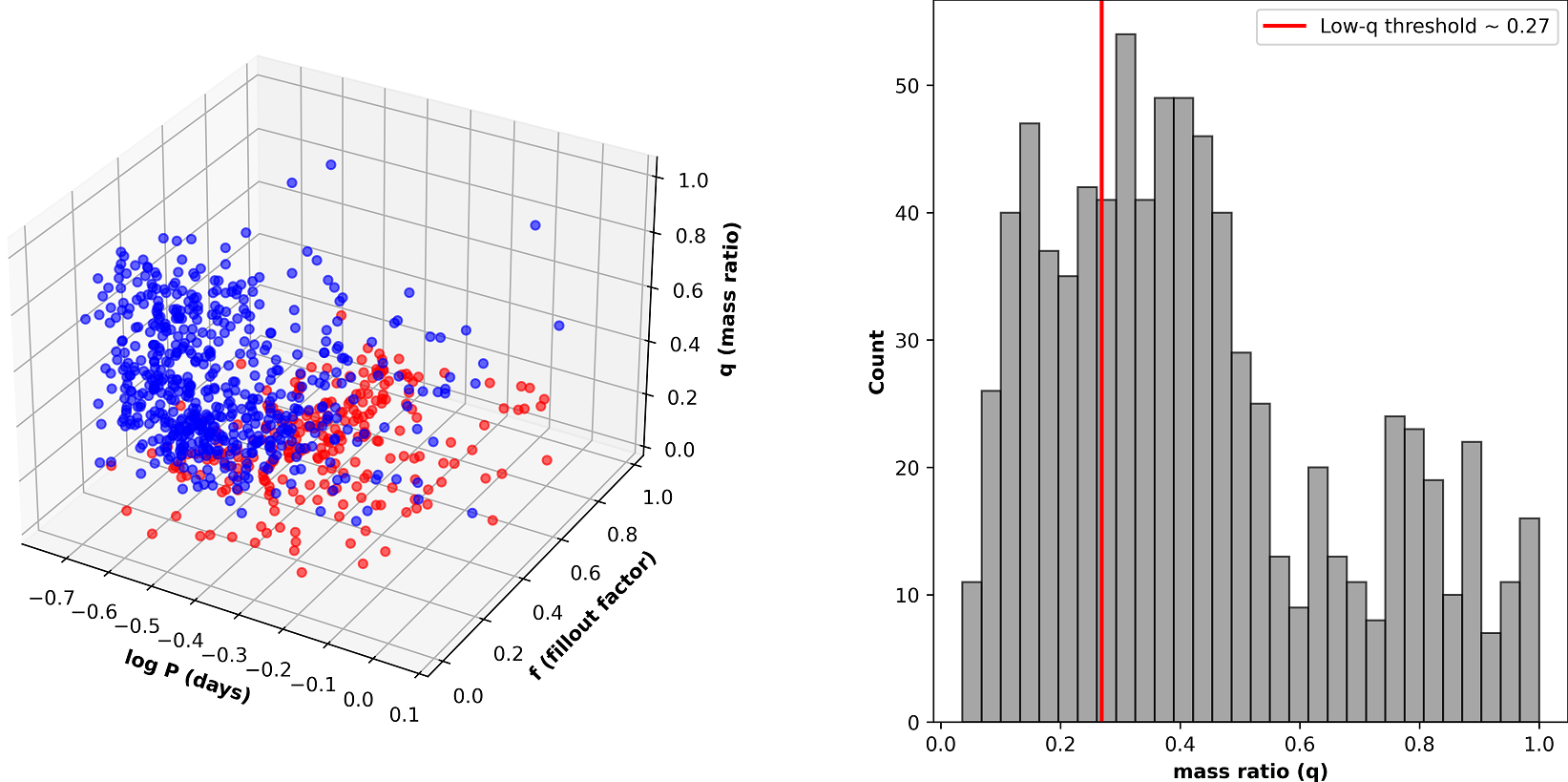}
\caption{Left: 3D scatter plot of $\log P$, $f$, and $q$ for the contact binary sample. Systems with $q \lesssim 0.27$ are highlighted in red. Right: histogram of $q$ with a vertical red line marking the low mass ratio threshold.}
\label{Fig:3Dscatter}
\end{figure*}

Figure \ref{Fig:3Dscatter} presents the results using two complementary visualizations. The left panel shows a 3D scatter plot of $P$, $f$, and $q$, with systems below the estimated low mass ratio threshold ($q \lesssim 0.27$) highlighted in red and the remaining systems in blue. The right panel displays the histogram of $q$ across the full sample, with a vertical red line marking the low mass ratio threshold. These visualizations clearly identify the low mass ratio systems while simultaneously conveying the distributions and interrelations of all three parameters in the sample.

This KDE-based approach thus provides a clear and reproducible way to explore the multidimensional parameter space of contact binaries. By simultaneously incorporating orbital period, fillout factor, and mass ratio, it defines an empirical upper limit for low mass ratio systems.
\\
\\
In the literature, contact binaries with extremely low mass ratios have attracted significant attention. However, their theoretical and observational cutoff in mass ratio remains highly challenging and not yet well established. Contact binary systems with extremely low mass ratios are studied from the perspective of theoretical dynamical stability (\citealt{1995ApJ...444L..41R, 2006MNRAS.369.2001L, 2007MNRAS.377.1635A, 2009MNRAS.394..501A, 2010MNRAS.405.2485J, 2015AJ....150...69Y, 2021MNRAS.501..229W, 2024SerAJ.208....1A, 2024MNRAS.527....1W}). Predicted minimum mass ratios have been refined over time, from early estimates of $q_{\rm min} \sim 0.09$ (\citealt{1995ApJ...444L..41R}) to more recent values as low as 0.038–0.041, depending on primary mass, stellar structural parameters, and metallicity (\citealt{2021MNRAS.501..229W, 2024NatSR..1413011Z, 2024SerAJ.208....1A, 2024MNRAS.527....1W}). Observationally, confirmed examples remain rare, including V1309 Sco (\citealt{2011A&A...528A.114T}), VSX J082700.8+462850 (\citealt{2021ApJ...922..122L}), TYC 4002-2628-1 (\citealt{2022MNRAS.517.1928G}), TYC 3801-1529-1 (\citealt{2024A&A...692L...4L}), and ASASSN-V J175200.35+361805.2 (\citealt{2025AJ....170..101G}), with mass ratios reaching as low as 0.027. These systems provide crucial observational benchmarks for studying the dynamics, stability, and merger potential of contact binaries with extremely low mass ratios (\citealt{2026ApJ...P}).

\section{Gyration and spin-orbit dynamics}
Contact binary stars are dynamically complex systems in which the interplay between stellar spin and orbital motion governs their long-term evolution. Understanding the balance between spin and orbital angular momentum is essential for assessing their rotational stability, particularly in the context of Darwin's instability criterion. In many studies, the dimensionless stellar structure constant $k_{1,2}$ (\citealt{2006MNRAS.369.2001L}), also referred to as the squared gyration radius, has commonly been adopted as $k_{1,2}=0.06$ for all contact binaries; however, this approximation is strictly valid only when both components are main sequence stars. For contact binaries with low mass ratios, this assumption may break down because the stellar components can differ significantly in their internal structure, leading to inaccurate stability assessments. Because the structural properties of the stellar components and their observational uncertainties strongly affect this balance, a statistical approach is required to properly account for parameter variations. This motivates a Monte Carlo (MC) analysis that simultaneously incorporates theoretical priors on internal structure and observational uncertainties, enabling a robust evaluation of the stability of contact binaries.

We performed a MC analysis to estimate the spin parameters and rotational stability of a sample of 501 contact binary stars collected from the \cite{2025MNRAS.538.1427P} study, where all systems have the required parameters available. The dataset included the primary and secondary masses, radii, semi-major axis, fillout factor, and squared gyration radii ($k_1^2, k_2^2$). Priors for $k_1^2$ and $k_2^2$ were adopted based on theoretical stellar structure models that provide typical values of the squared gyration radius for stars with different internal configurations. For the primary components, which are generally more massive and main sequence stars, we adopted $k_1^2 = 0.06(1)$, close to the classical values derived from polytropic models ($n=3$) and solar-type stellar structure calculations (\citealt{1993A&A...277..487C,2004A&A...424..919C}). For secondary components, the priors were adjusted according to their mass (or effective temperature) to reflect the stronger degree of central concentration or convection: low mass stars ($M_2 < 0.25,M_\odot$ or $T_\mathrm{eff} < 3800$ K), were assigned $k_2^2 = 0.18(3)$ (fully convective configuration, consistent with $n=1.5$ polytropes), intermediate-mass stars ($0.25 < M_2 < 0.6,M_\odot$) were given $k_2^2 = 0.10(2)$, and higher-mass secondaries were set to $k_2^2 = 0.06(1)$. These values are close to with stellar evolution calculations and widely adopted approximations in the literature (\citealt{1993A&A...277..487C,2004A&A...424..919C}).

For each of $N_\mathrm{MC} = 1000$ MC realizations, the squared gyration radii $k_1^2$ and $k_2^2$ were sampled from these priors and restricted to the physically reasonable interval $0.001 \leq k^2 \leq 0.4$, while the observed stellar parameters, mass $M_{1,2}$, radius $R_{1,2}$, and semi-major axis $a$ were taken directly from the dataset without perturbation. Spin and orbital angular momenta were then calculated as

\begin{equation}
J_\mathrm{spin} = \left( k_1^2 M_1 R_1^2 + k_2^2 M_2 R_2^2 \right)\,\Omega
\end{equation}

\begin{equation}
J_\mathrm{orb} = \mu a^2 \,\Omega
\end{equation}

\begin{equation}
\mu = \frac{M_1 M_2}{M_1 + M_2},
\end{equation}
and their ratio $J_\mathrm{spin}/J_\mathrm{orb}$ was evaluated.

To properly account for the consistency of the MC realizations with theoretical expectations, a log-likelihood function was computed for each realization:

\begin{equation}
\log \mathcal{L} = -\frac{1}{2} \left[ \frac{(k_1^2 - k_{1,\mathrm{th}}^2)^2}{\sigma_{k_1^2}^2} + \frac{(k_2^2 - k_{2,\mathrm{th}}^2)^2}{\sigma_{k_2^2}^2} \right] - \ln(\sigma_{k_1^2} \sigma_{k_2^2} 2\pi),
\end{equation}

\noindent where $k_{1,\mathrm{th}}^2$ and $k_{2,\mathrm{th}}^2$ are the theoretical squared gyration radii for the primary and secondary stars, respectively, and $\sigma_{k_1^2}, \sigma_{k_2^2}$ are the prior uncertainties. This log-likelihood quantifies the agreement between each MC sample and the theoretical priors, allowing the identification of the most probable realizations while naturally incorporating variations in stellar structure.

\begin{figure*}
\centering
\includegraphics[width=1\textwidth]{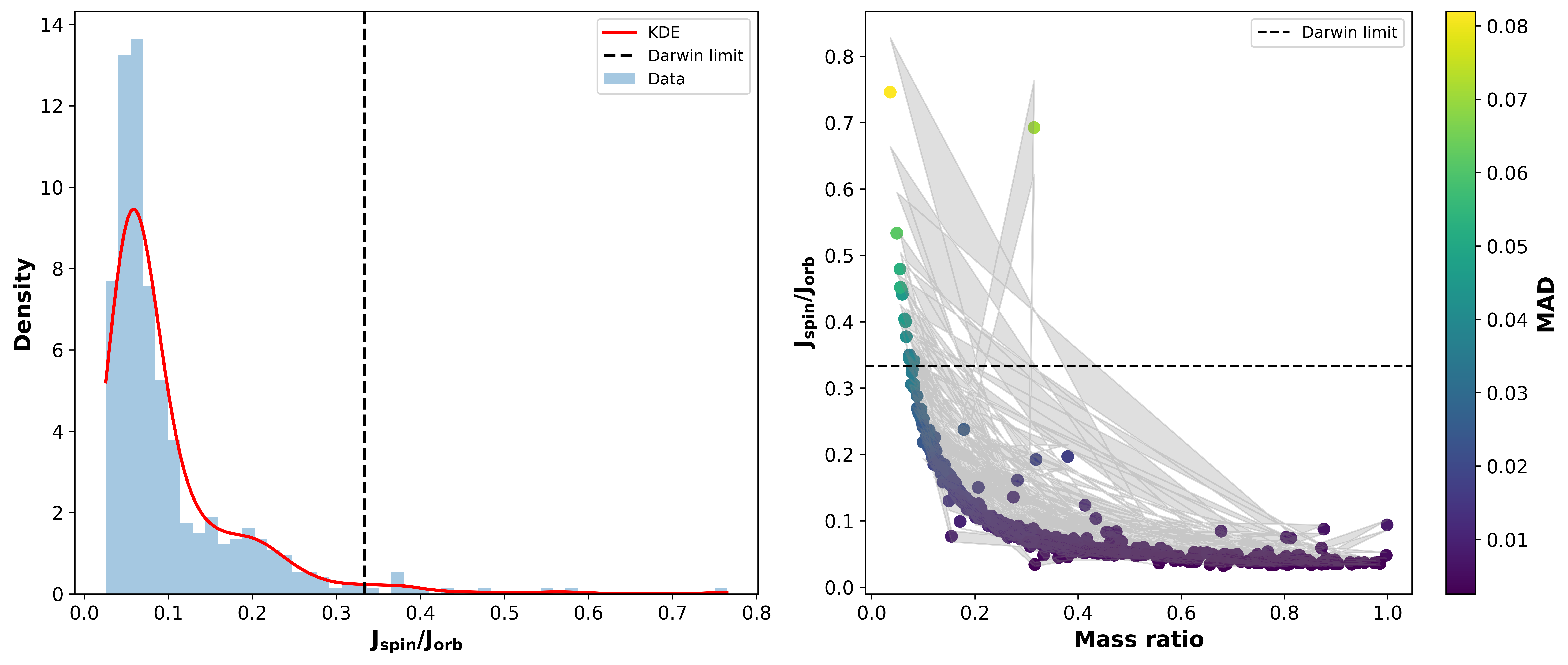}
\caption{a) Histogram of $J_\mathrm{spin}/J_\mathrm{orb}$ with a KDE overlay. The vertical line indicates the classical Darwin limit (0.333).
b) Two-dimensional scatter plot of the mass ratio versus $J_\mathrm{spin}/J_\mathrm{orb}$, with uncertainty bands represented by the MAD of MC realizations.}
\label{fig:k}
\end{figure*}

The mean and Median Absolute Deviation (MAD) of $k_1, k_2$, and $J_\mathrm{spin}/J_\mathrm{orb}$ were then derived over all realizations. Figure \ref{fig:k} presents a composite illustration of the results. Panel (a) shows the histogram of $J_\mathrm{spin}/J_\mathrm{orb}$ with a vertical line indicating the classical Darwin limit. Panel (b) presents the two-dimensional scatter plot of mass ratio versus $J_\mathrm{spin}/J_\mathrm{orb}$ with uncertainty bands represented by the MAD of MC realizations. These plots highlight the regions of stable and unstable systems across the sample.

\begin{figure}
\centering
\includegraphics[width=0.45\textwidth]{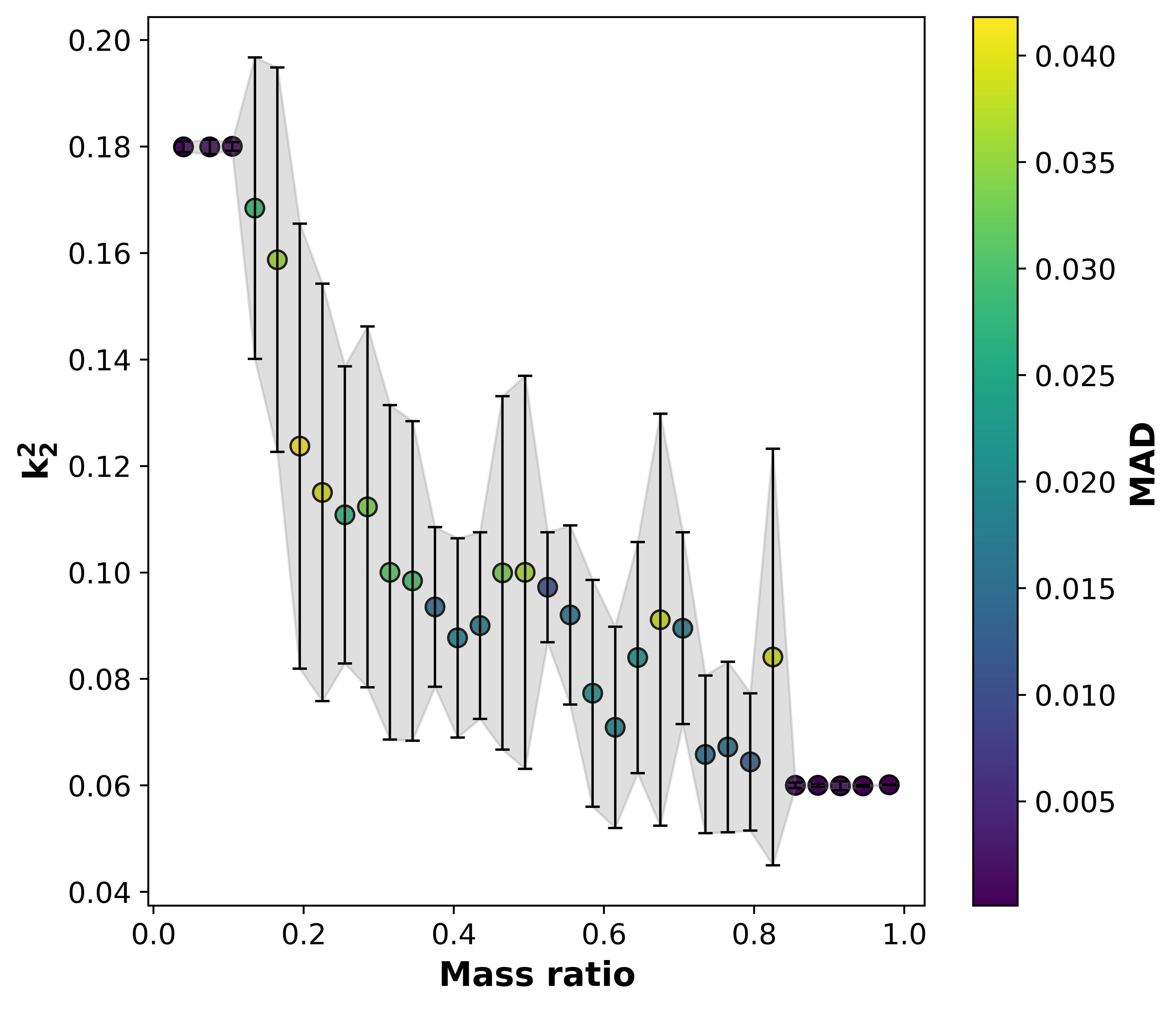}
\caption{Median values of $k_2^2$ versus mass ratio, computed over the mass ratio intervals listed in Table \ref{tab:q_bins}. The shaded regions denote the MAD, and colors encode the corresponding MAD.}
\label{fig:k2-q}
\end{figure}

Statistical summaries were computed by binning systems according to their mass ratio. Table \ref{tab:q_bins} presents the median and MAD of $k_1$, $k_2$, and $J_\mathrm{spin}/J_\mathrm{orb}$ within each $q$ bin. As seen, $k_1$ remains nearly constant, while $k_2$ and $J_\mathrm{spin}/J_\mathrm{orb}$ decrease slightly with increasing mass ratio, reflecting the influence of secondary structure on rotational stability (Figure \ref{fig:k2-q}).

\begin{table*}
\renewcommand\arraystretch{1.2}
\centering
\caption{Statistical summary of $k_1^2$, $k_2^2$, and $J_\mathrm{spin}/J_\mathrm{orb}$ for contact binaries, grouped by mass ratio bins. Columns: $q_\mathrm{bin}$ = mass ratio range, $k_{1,2}^2$ ± MAD, and $J_\mathrm{spin}/J_\mathrm{orb}$ ± MAD.}
\label{tab:q_bins}
\begin{tabular}{ccccccc}
\hline
$q_\mathrm{bin}$ & $k_1^2$ &  MAD & $k_2^2$ &  MAD & $J_\mathrm{spin}/J_\mathrm{orb}$ &  MAD \\
\hline
0.02--0.06 & 0.0599 & 0.0002 & 0.1799 & 0.0010 & 0.5177 & 0.1203 \\
0.06--0.09 & 0.0599 & 0.0003 & 0.1799 & 0.0013 & 0.3333 & 0.0413 \\
0.09--0.12 & 0.0600 & 0.0003 & 0.1800 & 0.0008 & 0.2313 & 0.0191 \\
0.12--0.15 & 0.0601 & 0.0002 & 0.1684 & 0.0283 & 0.1821 & 0.0176 \\
0.15--0.18 & 0.0601 & 0.0003 & 0.1587 & 0.0361 & 0.1473 & 0.0253 \\
0.18--0.21 & 0.0600 & 0.0003 & 0.1237 & 0.0418 & 0.1201 & 0.0101 \\
0.21--0.24 & 0.0599 & 0.0004 & 0.1150 & 0.0392 & 0.1016 & 0.0062 \\
0.24--0.27 & 0.0600 & 0.0002 & 0.1108 & 0.0279 & 0.0882 & 0.0059 \\
0.27--0.30 & 0.0600 & 0.0003 & 0.1123 & 0.0339 & 0.0841 & 0.0177 \\
0.30--0.33 & 0.0599 & 0.0002 & 0.1000 & 0.0314 & 0.0982 & 0.1156 \\
0.33--0.36 & 0.0599 & 0.0003 & 0.0984 & 0.0300 & 0.0673 & 0.0064 \\
0.36--0.39 & 0.0600 & 0.0003 & 0.0935 & 0.0150 & 0.0662 & 0.0248 \\
0.39--0.42 & 0.0600 & 0.0003 & 0.0877 & 0.0187 & 0.0620 & 0.0138 \\
0.42--0.45 & 0.0601 & 0.0004 & 0.0900 & 0.0175 & 0.0579 & 0.0092 \\
0.45--0.48 & 0.0601 & 0.0002 & 0.0999 & 0.0332 & 0.0581 & 0.0102 \\
0.48--0.51 & 0.0599 & 0.0002 & 0.1000 & 0.0369 & 0.0548 & 0.0056 \\
0.51--0.54 & 0.0601 & 0.0002 & 0.0972 & 0.0103 & 0.0533 & 0.0037 \\
0.54--0.57 & 0.0600 & 0.0004 & 0.0920 & 0.0168 & 0.0491 & 0.0062 \\
0.57--0.60 & 0.0598 & 0.0004 & 0.0773 & 0.0213 & 0.0452 & 0.0049 \\
0.60--0.63 & 0.0601 & 0.0003 & 0.0709 & 0.0189 & 0.0418 & 0.0046 \\
0.63--0.66 & 0.0600 & 0.0003 & 0.0840 & 0.0217 & 0.0467 & 0.0073 \\
0.66--0.69 & 0.0600 & 0.0003 & 0.0911 & 0.0387 & 0.0466 & 0.0156 \\
0.69--0.72 & 0.0600 & 0.0005 & 0.0895 & 0.0180 & 0.0460 & 0.0036 \\
0.72--0.75 & 0.0598 & 0.0003 & 0.0658 & 0.0148 & 0.0406 & 0.0046 \\
0.75--0.78 & 0.0600 & 0.0003 & 0.0672 & 0.0160 & 0.0397 & 0.0059 \\
0.78--0.81 & 0.0600 & 0.0003 & 0.0644 & 0.0129 & 0.0420 & 0.0130 \\
0.81--0.84 & 0.0602 & 0.0004 & 0.0841 & 0.0391 & 0.0434 & 0.0118 \\
0.84--0.87 & 0.0601 & 0.0002 & 0.0600 & 0.0005 & 0.0368 & 0.0030 \\
0.87--0.90 & 0.0600 & 0.0002 & 0.0600 & 0.0003 & 0.0438 & 0.0162 \\
0.90--0.93 & 0.0602 & 0.0001 & 0.0599 & 0.0008 & 0.0359 & 0.0017 \\
0.93--0.96 & 0.0601 & 0.0003 & 0.0599 & 0.0002 & 0.0361 & 0.0013 \\
0.96--1.00 & 0.0600 & 0.0002 & 0.0601 & 0.0001 & 0.0367 & 0.0010 \\
\hline
\end{tabular}
\end{table*}

\section{Empirical parameter relationships}
\subsection{New dataset for low mass ratio contact systems}
We compiled a dataset from the literature on contact binary systems with low mass ratios, defined as $q < 0.27$ according to the threshold established in Section 2. For each system, we gathered the parameters reported in the literature, including orbital period ($P$), orbital inclination ($i$), fillout factor ($f$), $T_{1,2}$, $r_{1,2}$ and $l_{1,2}$, as well as reports on the presence of starspots or third-body companions. Absolute parameters published in previous studies were also included when available. Subsequently, we recalculated the absolute parameters for all systems using Gaia DR3 parallaxes. To ensure the consistency and reliability of the sample, several exclusion criteria were applied. Systems without Gaia DR3 parallaxes, with interstellar extinction ($A_V$) $> 0.4$ (\citealt{2024NewA..11002227P}), or lacking essential input parameters for the calculations (component effective temperatures, $q$, $r_{1,2}$, and $l_{1,2}$ from published sources; maximum brightness ($V_\mathrm{max}$) from the VSX or AAVSO database; and orbital period $P$ from ASAS-SN) were excluded. The present sample comprises contact binary systems with orbital periods shorter than 0.6 days, in accordance with the criteria adopted by \cite{2022MNRAS.510.5315P}. In addition, systems showing discrepancies larger than 0.1 between semi-major axes, which may indicate unreliable solutions, were excluded following the criterion of \cite{2024NewA..11002227P}. After applying these filtering steps, the final dataset consisted of 115 low mass ratio contact binaries. This sample is made available in a machine-readable format in the online version of the paper.

The absolute visual magnitude ($M_V$) of each system was calculated from the $V_\mathrm{max}$, the distance from Gaia DR3 parallaxes ($d$), and $A_V$ (Equation \ref{eqMv}). The magnitudes of the individual components ($M_{V1}$ and $M_{V2}$) were obtained using the luminosity ratios ($l_{1,2}/l_\mathrm{tot}$) (Equation \ref{eqMv1,2}).

\begin{equation}\label{eqMv}
M_{V(system)}=(V_{max})-5log(d)+5-(A_V),
\end{equation}

\begin{equation}\label{eqMv1,2}
M_{V(1,2)}-M_{V(system)}=-2.5log(\frac{l_{(1,2)}}{l_{(tot)}}).
\end{equation}

Bolometric magnitudes ($M_{\text{bol1,2}}$) were computed by applying bolometric corrections ($BC_1,2$), and stellar luminosities ($L_{1,2}$) were then derived from these bolometric magnitudes. Stellar radii ($R_1$ and $R_2$) were determined from the $L_{1,2}$ and effective temperatures ($T_{1,2}$).  

The semi-major axis of each system was estimated from the stellar radii (${R_1,2}$) and $r_{1,2}$; two values, $a_1$ and $a_2$, were calculated independently, and the mean was adopted when the difference between them ($\Delta a$) was smaller than 0.1. Using the semi-major axis together with the orbital period and mass ratio, the stellar masses were calculated according to Kepler’s third law:

\begin{equation}
M_1 = \frac{4\pi^2 a^3}{G P^2 (1+q)}, \qquad M_2 = q \times M_1.
\end{equation}

Surface gravities ($\log g_1$ and $\log g_2$) were then derived from the $M_{1,2}$ and $R_{1,2}$). The orbital angular momentum of each system was computed as:

\begin{equation}\label{eqJ0}
J_{\mathrm{orb}} \equiv J_0 =
\frac{q}{(1+q)^2} \sqrt[3]{\frac{G^2}{2\pi} M_{\mathrm{tot}}^5 P},
\end{equation}
where $M_{tot}$ is the total mass of the system.

\subsection{Relationships}
Although numerous theoretical and observational studies have been devoted to contact binary stars, our understanding of their physical behaviors remains incomplete. Nevertheless, continuous revisions of empirical analyses concerning the parameter relations of these systems can be highly beneficial. The available results on such relations reveal both strong and weak trends, depending on the selected sample and the methodology employed. Therefore, sustained efforts to revisited these relationships using larger and more comprehensive samples are of considerable importance, particularly since such relations can serve as valuable tools for estimating the absolute parameters of these systems. Such empirical revisions of parameter relations are usually carried out for the entire class of contact binary systems, sometimes with additional constraints. For instance, \cite{2021ApJS..254...10L} applied an orbital period limit of 0.5 days in his updates of empirical parameters, motivated by the presence of a break in the period–temperature diagram. Similarly, other studies, such as \cite{2022MNRAS.510.5315P}, adopted a period limit of 0.6 days. In some cases, these revisions have also been restricted to the A- and W-subtypes of contact binaries. However, no empirical relations or dedicated samples have yet been provided for the subclass of contact binaries with low mass ratios. As presented in Section 2, this type of contact binaries can exhibit certain common relative properties. Considering that in these systems both the mass and radius ratios differ from those of other contact binaries, we have attempted to independently examine and provide empirical relationships for their parameters.

Using the constructed sample of low mass ratio contact binaries, for which absolute parameters were estimated from Gaia DR3 parallaxes, we provide independent empirical relations for the parameters of these systems. The empirical relationships for low mass ratio contact binaries considered in this study include the following:  $P\!-\!M_{1,2}$, $P\!-\!R_{1,2}$, $P\!-\!L_{1,2}$, $P\!-\!M_{\rm bol\,1,2}$, $P\!-\!\log g_{1,2}$, $P\!-\!a$, $M_{1,2}\!-\!L_{1,2}$, and $M_{1,2}\!-\!R_{1,2}$. All of these relationships have been derived and are presented in Table \ref{relations}.

\begin{table}
\renewcommand\arraystretch{1.2}
\centering
\caption{Extracted empirical relationships of parameters with uncertainties.}
\begin{tabular}{c c}
\hline
Relation & Equation \\
\hline
$P-a$ & $a = (6.189 \pm 0.562)\,P + (0.200 \pm 0.222)$ \\
$P-M_1$ & $M_1 = (2.835 \pm 0.782)\,P + (0.357 \pm 0.310)$ \\
$P-M_2$ & $M_2 = (0.409 \pm 0.161)\,P + (0.076 \pm 0.064)$ \\
$P-R_1$ & $R_1 = (4.037 \pm 0.365)\,P + (-0.149 \pm 0.145)$ \\
$P-R_2$ & $R_2 = (0.864 \pm 0.265)\,P + (0.365 \pm 0.105)$ \\
$P-L_1$ & $L_1 = (23.174 \pm 1.491)\,P + (-5.630 \pm 0.590)$ \\
$P-L_2$ & $L_2 = (3.451 \pm 0.352)\,P + (-0.723 \pm 0.139)$ \\
$P-\log g_1$ & $\log g_1 = (-1.197 \pm 0.154)\,P + (4.711 \pm 0.061)$ \\
$P-\log g_2$ & $\log g_2 = (-1.190 \pm 0.151)\,P + (4.578 \pm 0.060)$ \\
$P-M_{\rm bol1}$ & $M_{\rm bol1} = (-7.217 \pm 0.722)\,P + (8.300 \pm 0.286)$ \\
$P-M_{\rm bol2}$ & $M_{\rm bol2} = (-8.814 \pm 0.662)\,P + (7.100 \pm 0.262)$ \\
$M_1-L_1$ & $L_1 = (1.800 \pm 0.249)\,M_1 + (0.759 \pm 0.397)$ \\
$M_2-L_2$ & $L_2 = (1.868 \pm 0.207)\,M_2 + (0.180 \pm 0.056)$ \\
$M_1-R_1$ & $R_1 = (0.478 \pm 0.040)\,M_1 + (0.724 \pm 0.064)$ \\
$M_2-R_2$ & $R_2 = (1.207 \pm 0.109)\,M_2 + (0.418 \pm 0.029)$ \\
\hline
\end{tabular}
\label{relations}
\end{table}

\begin{figure*}[ht!]
\centering
\includegraphics[width=0.89\textwidth]{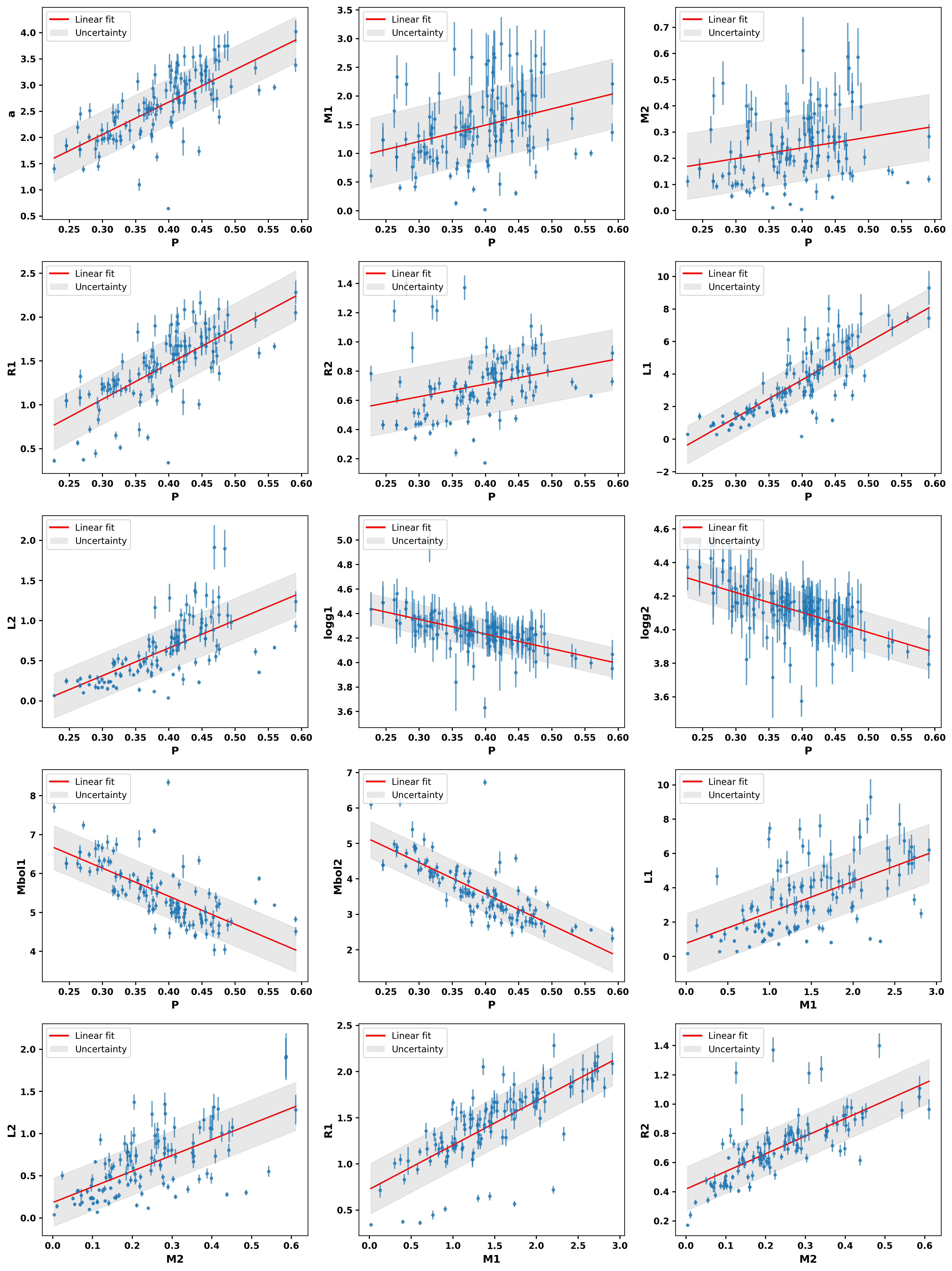}
\caption{Scatter plots illustrating correlations between selected stellar parameters in low mass ratio contact binaries with orbital periods shorter than 0.6 days. Linear regression lines are overlaid: red lines correspond to fits using mean errors, and blue lines correspond to fits using literature-reported errors. Shaded areas indicate the $\pm 1\sigma$ uncertainty ranges for each fit.}
\label{fig:rel}
\end{figure*}

Linear regressions were performed to investigate correlations among the derived absolute parameters of contact binaries, using a single approach based on a representative average uncertainty of the derived parameters. Fits were obtained via standard least-squares minimization, assuming the same characteristic uncertainty for all data points. The results are presented in Figure \ref{fig:rel}, where the regression solutions are shown with their $\pm 1\sigma$ confidence ranges. These two-dimensional relations are presented using empirical parameters in solar units. Furthermore, as illustrated in Figure \ref{fig:rel}, they can be satisfactorily described by linear fits.

\vspace{0.6cm}
\section{Discussion and Conclusion}
This study addresses several aspects of a specific group of contact binaries, which in the literature are commonly referred to as low mass ratio systems. The first objective was to establish an empirical upper boundary for the mass ratio in this group, and the possible cutoff was also discussed. In addition, results from a MC analysis of spin-orbit stability were presented, providing insights that may contribute to future investigations. By compiling and estimating the absolute parameters, we constructed a dedicated sample of low mass ratio contact systems, which enabled us to examine and propose empirical relations among their physical parameters. The discussion and conclusions regarding the findings are presented as follows:

A) Previous studies of W UMa-type contact binaries did not identify a clear boundary for low mass ratio systems, and as a result, the range of physical parameters defining such systems remained uncertain. In this work, we analyzed a large sample of contact binaries described in Section 2 and identified an empirical upper boundary for low mass ratio systems based on their physical properties. By examining independent parameters such as orbital period, fillout factor, inclination, and component temperatures, we aimed to identify the range of mass ratios corresponding to low mass ratio systems and to investigate which physical characteristics most clearly distinguish them from higher mass ratio binaries. Correlation analysis shows that while no single physical parameter alone provides a complete separation of low and higher mass ratio systems, certain parameters, particularly the fillout factor and orbital period, contribute most strongly to distinguishing the populations when considered together. Systems with mass ratios below approximately 0.27 tend to display significantly higher fillout factors, which can be explained by the weaker gravitational influence of the secondary star that allows the primary envelope to expand further and form a more developed common envelope. These binaries also generally have shorter orbital periods, consistent with the requirement of closer separations under strong overcontact conditions, while component temperatures show only minor differences compared with higher mass ratio systems.

B) Contact binary stars exhibit complex interactions between stellar spin and orbital motion, and stability assessments based on a fixed structural constant may be inaccurate for low mass ratio systems due to differences in stellar internal structure. To address this, we sampled the squared gyration radii $k_1^2$ and $k_2^2$ from theoretical priors while keeping the observed stellar parameters fixed, and calculated $J_\mathrm{spin}/J_\mathrm{orb}$. The results show that $k_1$ remains nearly constant, while $k_2$ and $J_\mathrm{spin}/J_\mathrm{orb}$ decrease slightly with increasing mass ratio, indicating that the secondary's internal structure influences rotational stability. The statistical summary provided in Table \ref{tab:q_bins} offers a reference for future studies aiming to model or analyze the rotational properties of low mass ratio contact binaries.

C) We compiled a dataset of 115 low mass ratio ($q < 0.27$) contact binary systems, collecting orbital and stellar parameters from the literature and calculating absolute values using Gaia DR3 parallaxes. To ensure reliability, we retained only systems with orbital periods shorter than 0.6 days and excluded those lacking essential data, with high extinction ($A_V > 0.4$), or showing discrepancies larger than 0.1 between $a_1$ and $a_2$. Then, using this sample, we derived independent empirical parameter relations for low mass ratio systems. These relations include orbital period versus mass, radius, luminosity, absolute bolometric magnitude, surface gravity, and semi-major axis, as well as mass–luminosity ($M-L$) and mass–radius ($M-R$) relations for each component. These empirical relationships can serve as a practical tool for estimating the absolute parameters of low mass ratio contact binaries in cases where high-resolution spectroscopy of both components is not available.

It is important to account for the uncertainties in the measured quantities when performing regression analyses. Simple regressions that consider only the nominal values can yield significantly biased results, particularly when a small number of highly precise measurements dominate over a larger number of less precise data points. However, the uncertainties in the absolute parameters for the 115 systems in the sample depend on the uncertainties of the light curve solutions collected from the literature. These light curve solutions are derived using various methods in different studies, and their levels of precision can vary. Therefore, we used the mean uncertainties of the derived parameters. As shown in Figure \ref{fig:rel}, some apparently outlying points are also visible. These may arise for different reasons: a system may possess unique characteristics that distinguish it from the rest of the sample, or the result may stem from suboptimal solutions in the literature. In any case, although such points can be treated as outliers in statistical analyses, they should not be physically disregarded.

\begin{figure*}[ht!]
\centering
\includegraphics[width=1\textwidth]{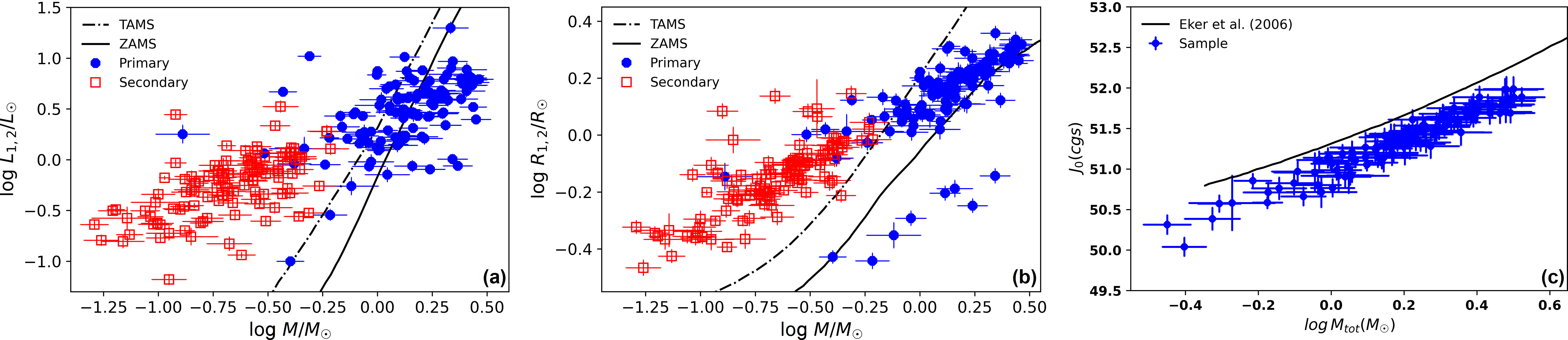}
\caption{a) $M_{1,2}-L_{1,2}$, b) $M_{1,2}-R_{1,2}$, c) $M_{\mathrm{tot}}$–$J_0$. The diagrams are presented using a logarithmic scale.}
\label{fig:MLRJ0}
\end{figure*}

\begin{figure}[ht!]
\centering
\includegraphics[width=0.5\textwidth]{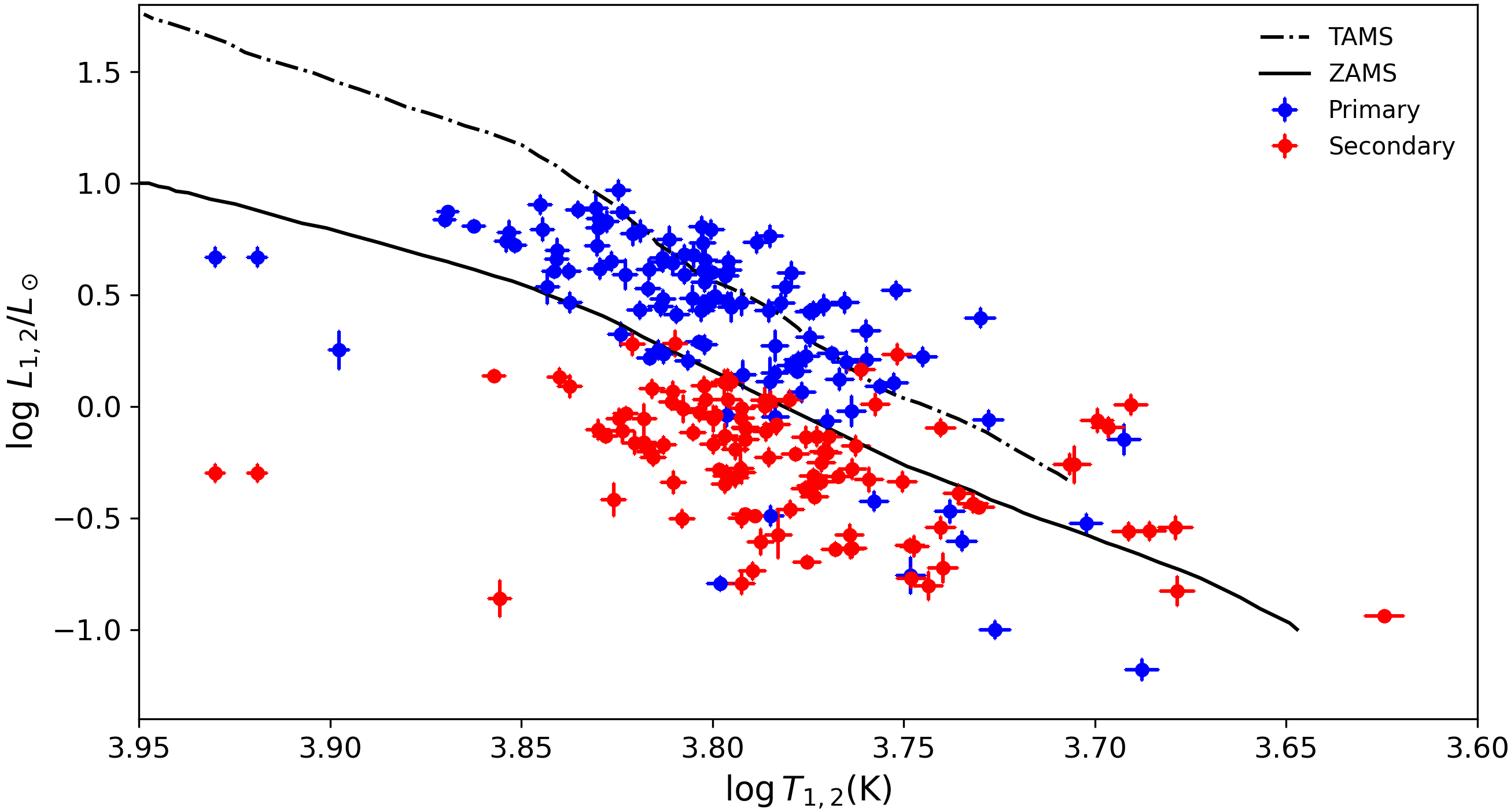}
\caption{H-R diagram showing the positions of the primary and secondary stars in 115 low mass ratio binary systems, illustrating their distribution relative to the ZAMS and TAMS lines.}
\label{fig:HR}
\end{figure}

The estimated absolute parameters allow us to place the stellar components on the $M-L$ and $M-R$ diagrams (Figure \ref{fig:MLRJ0}a,b). Figure \ref{fig:MLRJ0}a,b shows the components in comparison with the Zero-Age Main Sequence (ZAMS) and Terminal-Age Main Sequence (TAMS) boundaries as established by \cite{2000AAS..141..371G}. As shown in Figure \ref{fig:MLRJ0}a,b, the secondary stars are located above the TAMS, while the more massive primary stars lie near the ZAMS. In some systems, however, the primary stars are also positioned above the TAMS, like their secondary companions, indicating that both stars have evolved. Also, Figure \ref{fig:MLRJ0}a shows that a significant number of primary stars are located below the ZAMS. It should be emphasized that contact binaries undergo complex evolutionary processes and interactions (\citealt{2005ApJ...629.1055Y}, \citealt{2011AcA....61..139S}), which cause their evolutionary stages to differ substantially from those of single stars. Therefore, any comparison with the standard single-star ZAMS and TAMS lines should be interpreted cautiously. As shown in Figure \ref{fig:HR}, the distribution of the secondary components emphasizes that caution is required when interpreting their evolutionary status. A significant fraction of the secondaries lie below the Hertzsprung-Russell (H-R) diagram, indicating lower luminosities, while their positions in the $M$-$L$ and $M$-$R$ diagrams are influenced by Roche lobe filling and energy transfer (Figure \ref{fig:MLRJ0}a,b).

The orbital angular momentum of each system, estimated using Equation \ref{eqJ0}, is shown in the $\log M_{\mathrm{tot}}$–$J_0$ diagram for the sample systems (Figure \ref{fig:MLRJ0}c). The area below the quadratic line in Figure \ref{fig:MLRJ0}c is generally associated with contact binary stars, whereas the area above corresponds to detached systems (\citealt{2006MNRAS.373.1483E}).

\section*{Data Availability}
The new sample used in this study is provided in machine-readable format within the online version of the paper.

\begin{acknowledgements}
This work was carried out within the framework of the BSN project(\url{https://bsnp.info}). We used data from the European Space Agency mission Gaia(\url{http://www.cosmos.esa.int/gaia}). We are grateful to Ehsan Paki and Ghazal Alizadeh for their valuable assistance. We would like to sincerely thank David Valls-Gabaud for his valuable feedback and guidance, which greatly contributed to the improvement of this manuscript. We would like to acknowledge the assistance of Elham Sarvari.
\end{acknowledgements}


\end{document}